\documentclass[aps,pra,twocolumn,superscriptaddress,floatfix]{revtex4-1}
\usepackage{amssymb}
\usepackage{amsmath}
\usepackage{wasysym}
\usepackage{verbatim}
\usepackage{enumerate}
\usepackage[colorlinks,bookmarks=false,citecolor=blue,linkcolor=red,urlcolor=blue]{hyperref}
\usepackage[usenames,dvipsnames]{color}
\usepackage{times}
\usepackage{color}
\usepackage{graphicx}
\usepackage{epsfig}
\usepackage{float}
\usepackage{bbold}
\usepackage{tabularx}
\usepackage{natbib}
\usepackage{pdfpages}

\pdfoutput=1
\newcommand {\ket}[1] {|#1 \rangle}
\newcommand {\bra}[1] {\langle#1 |}
\newcommand {\braket}[2] {\langle #1 | #2 \rangle}
\newcommand {\ketbra}[2] {| #1 \rangle \langle #2 |}

\newcommand{\Eref}[1]{Eq. (\ref{#1})}

\newcommand{\Fref}[1]{Fig. \ref{#1}}
\newcommand{\Sref}[1]{Sec. \ref{#1}}

\newcommand{\etal}{{\it et al.}~}

\newcommand{\be}{\begin{equation}}
\newcommand{\ee}{\end{equation}}
\newcommand{\bea}{\begin{eqnarray}}
\newcommand{\eea}{\end{eqnarray}}

\newcommand{\Smax}{S_\mathrm{max}}
\newcommand{\Savg}{S_\mathrm{avg}}

\begin{document}

\title{Superoperators vs. Trajectories for Matrix Product State Simulations of Open Quantum System: A Case Study}

\author{Lars Bonnes}
\email{lars.bonnes@uibk.ac.at}
\author{Andreas M. L\"auchli}
\address{Institute for Theoretical Physics, University of Innsbruck, Technikerstr. 25, A-6020 Innsbruck, Austria}

\date{\today}

\begin{abstract}
Quantum trajectories and superoperator algorithms implemented within the matrix product state (MPS) framework are powerful tools to simulate the real-time dynamics of open dissipative quantum systems.
As for the unitary case, the reachable time-scales as well as system sizes are limited by the (possible) build-up of entanglement entropy.
The aforementioned methods constitute complementary approaches how Lindblad master equations can be integrated relying either on a quasi-exact representation of the full density matrix or a stochastic unraveling of the density matrix in terms of pure states.
In this work, we systematically benchmark both methods by studying the dynamics of a Bose-Hubbard chain in the presence of local as well as global dephasing. 
The build-up as well as system-size scaling of entanglement entropy strongly depends on the method and the parameter regime and we discuss the applicability of the methods for these cases as well as study the distribution of observables and time discretization errors that can become a limiting factor for global dissipation.
\end{abstract}
\pacs{}
\maketitle

\section{Introduction}
The field of open, driven and dissipative quantum system has gained renewed interest from the theoretical and experimental point of view over the last few years.
The typical physical setup consists of a quantum system $\mathcal{A}$ that is coupled to an environment $\mathcal{B}$.
By tracing out the environmental degrees of freedom an equation of motion for the density matrix $\rho_\mathcal{A}$ of system $\mathcal{A}$ can be obtained~\cite{breuer10}.
Although these concepts are well established in quantum optics -- for instance driven cavity systems that can also feature dissipative phases such as self-organized Dicke supersolids~\cite{baumann10,baumann11} and glass-like phonemena~\cite{buchhold13} --, in exciton-polariton condensates~\cite{kasprazak06,carusotto13} or as unwanted sources of noise and heating due to stray light fields~\cite{pichler10} or three-body collisions, ideas emerged to utilize tailored baths in quantum many-body systems to engineer pure and non-trivial quantum states using cold atoms or trapped ions~\cite{diehl08,kraus08,diehl11b,bardyn12,bardyn13,stannigel13}.
These advances raise new questions about dynamic critical phenomena ~\cite{sieberer13} and dynamical phase transitions~\cite{tomadin11} in the presence of dephasing.

In light of these recent developments it is imperative to provide numerical techniques that are able to approach of these problems.
In one dimension matrix product states (MPS) provides an efficient framework to represent quantum states and methods such  as the density matrix renormlization group (DMRG)~\cite{white92,white93,peschel99,schollwoeck05,schollwoeck10} have been extended such that even real-time dynamics of mixed (thermal) states can be accessed~\cite{feiguin05,sirker05,karrasch12,barthel13,karrasch13,bonnes14c}.

It is also possible to extend the ideas developed in the context of unitary (Schr\"odinger) dynamics to Lindblad quantum master equations (QME)~\cite{kossakowski72,gorini76,lindblad76,gardinger04,breuer10} of the form
\begin{equation}
 \partial_t \rho = -i [ \mathcal{H}, \rho ]
 + \sum_{i} \kappa_\alpha \left( c_\alpha \rho c^\dagger_\alpha - \frac{1}{2}\{ c^\dagger_\alpha c_\alpha, \rho \} \right).
   \label{eq:master}
\end{equation}
Here, $\mathcal{H}$ is the Hamiltonian of the one-dimensional lattice system and the $c_i$'s are time-local jump operators.
One approach, developed in the context of quantum optics, is a stochastic method to integrate \Eref{eq:master}.
The \textit{quantum trajectory} method~\cite{dum92,gisin92,dalibard92,moelmer93,breuer97,plenio98,breuer10,daley14} unravels the full master equation into an ensemble of pure states that evolves according to an effective Hamiltonian and is supplemented by random ``quantum jump'' events.
The density matrix is recovered by averaging over the ensemble of quantum states.
This method only operates on pure quantum states and allowed for a study of much larger systems as the memory requirenments for the direct integration of the equations of motion (using Runge-Kutta, for instance) scales quadratically in the Hilbert space size for the density matrix but only linearly for a state.
In fact, the computational performance of the stochastic approach has been demonstrated to be superior to a direct integration of the full density matrix for specific problems~\cite{breuer97}.
Daley \etal~\cite{daley09} implemented quantum trajectories using time dependent DMRG and applied it to interacting quantum lattice models.
Since then, it has been applied to different dissipative systems such as three-body losses in optical lattices~\cite{daley09,kantian09}, dissipative defects~\cite{barmettler11} or the short-time dynamics of dephasing~\cite{pichler10}.

A complementary approach is the representation of the density matrix as a matrix product operator~\cite{verstraete04a} or as a purified state state evolving according to a superoperator Hamiltonian~\cite{zwolak04,prosen09}.
The dynamics if fully deterministic, i.e. it does not involve statistical sampling, on the cost of operating on a larger Hilbert space.
The super state can be updated using standard MPS methods and has proven useful in the study of operator space entanglement~\cite{pizorn09}, non-equilibrium spin systems~\cite{prosen09,znidaric10,vannieuwenburg14}, and engineered dissipation~\cite{bonnes14a}.

Although MPS approaches to open system dynamics have been applied to a wide class a problems, a systematic comparison of the performance of the stochastic approach and the superoperator technique has only been performed for  sparse-matrix implementations in quantum optics -- see the comparative study in Ref.~\cite{breuer97}.
The efficiency of the MPS approach, however, is highly sensitive towards the amount of entanglement entropy generated during the time evolution as the matrix dimension used in the MPS representation scales approximately exponentially with the amount of entanglement encoded in each link.
It is thus timely to reevaluate the previous results~\cite{breuer97} in light of the particular class of states.

In this paper, we compare the quantum trajectory algorithm with the superoperator renormalization scheme implemented using the time-evolving block decimation (TEBD) algorithm~\cite{vidal03,vidal04}.
The bechmark system is a Bose-Hubbard chain subject to local or global dephasing.
The build-up of operator space entanglement and the distribution of entanglement entropy in the trajectory ensemble gives us a direct measure to quantify the complexity of the problem in one or the other algorithm and we find that these properties strongly depend on the locality and the strength of the dissipation making this dissipative Hubbard chain a good benchmark system.
Although the steady-state of this process is always the infinitely-hot maximally mixed state, we find that local dissipation can even lead to an extensive steady-state entanglement in the trajectories compared to a logarithmic divergence in the superoperator case.
This highlights that the applicability of both methods can strongly depend on the parameters of the dissipation.
We also analyze the variance for some observables in the trajectory ensemble and provide important bounds for the Trotter time step due to extensively small jump times.

This manuscript is organized as follows.
In \Sref{sec:res} we introduce the Bose-Hubbard chain and the dephasing mechanism considered and briefly review the numerical algorithms as well as the entanglement measures.
Our main results for local and global dissipation are presented in \Sref{sec:res} and finally summed up and discussed in \Sref{sec:discussion} and \ref{sec:conclusion}.

\section{Model and Methods}

\subsection{Dephasing in the Bose-Hubbard Chain}
\label{sec:model}

The model considered here is a Bose-Hubbard chain of length $L$ with on-site interaction $U$ and nearest-neighbor hopping $t$,
\begin{equation}
 \mathcal{H}=-t \sum_{i=1}^{L-1} \left( b_i^\dagger b_j + b_j^\dagger b_i \right) 
	+ \frac{U}{2} \sum_i n_i (n_i -1), 
\end{equation}
with $b_i$, $b^{\dagger}_i$ being the standard bosonic annihilation and creation operators and $n_i = b_i^\dagger b_i$.
At integer fillings it exhibits a phase transition from a gapless superfluid state to a Mott insulator~\cite{kuehner98,kuehner00,giamarchi04}.
It is highly generic as it is non-integrable except in the $U=0$ and the hard-core limit and is widely used to describe, for instance, bosonic atoms in the lowest band of an optical lattice.
The local Hilbert space is truncated at $d=n_\mathrm{max}+1$ where $n_\mathrm{max}$ is the maximally allowed number of bosons per lattice site.

In addition to the unitary dynamics generated by $\mathcal{H}$, a mechanism of dephasing is incorporated in a Lindblad quantum master equation as given in \Eref{eq:master}.
The Lindblad operators are given by $c_i=n_i$, i.e. the bath performs local density measurements.
Physically, this master equation can describe heating from incoherent light scattering in optical lattices~\cite{pichler10,poletti12,bonnes14c}.
Thus, the system is eventually driven into the infinitely hot state that is stationary under the time evolution in \Eref{eq:master}. 

This model is a suitable benchmark system as local and global dissipation -- the dissipation acts either on one site locally, e.g. $\kappa_i =0$ for $i \ne L/2$, or on all sites will equal coupling ($\kappa_i = \kappa$ for all $i$) -- show qualitatively different behaviors in terms of time scales and growth of operator space entanglement in the transient regime.
In particular, the operator space entropy exhibits a volume law scaling~\cite{bonnes14c} for local dissipation and is thus comuptationally very hard to treat whereas a logarithmic scaling is observed in the global case.
It is yet unclear, how this behavior will reflect in the quantum trajectory approach and whether it is possible to access larger system sizes and/or longer times using a stochastic approach.

\subsection{Time Evolving Block Decimation}
The real time evolution for one-dimensional quantum systems with short range -- here nearest neighbor -- interactions, $L$ sites and open boundary conditions can effectively be simulated using the TEBD algorithm introduced by Vidal~\cite{vidal03,vidal04} and is one of the standard techniques applied to numerous physical systems.
The starting point is the MPS representation of the initial state $\ket \Psi$ using an orthonormal basis $\lbrace \ket{i_1 i_2 ... i_L} \rbrace$ reading
\begin{equation}
\ket{\Psi} = \sum_{i_1, i_2, ..., i_L} \mathbf{B}_1^{i_1} \mathbf{\Lambda}_2 \mathbf{B}_2^{i_2} ... \mathbf{\Lambda}_{L} \mathbf{B}^{i_L}_L \ket{i_1 i_2 ... i_L},
\label{eq:MPS}
\end{equation}
where $\mathbf{\Lambda}_n=\mathrm{diag}(\lambda_1, ..., \lambda_\chi)$ are diagonal matrices containing the (truncated) Schmidt coefficients obtained from consecutive singular value decompsitions of the state matrix.
The $\mathbf{B}_n^{i_n}$ are a set of right-orthonormal matrices
 of size $\chi \times \chi$ for each physical index $i_n$.
$\ket \Psi$ is evolved in time according the Schr\"odinger equation by application of a unitary gate $U_t=\exp(-i \mathcal H t)$, i.e. $\ket{\Psi(t)} = U_t \ket \Psi$.
The nearest-neighbor nature of the interaction allows us to write $\mathcal{H} = \mathcal{H}_\mathrm{e} + \mathcal{H}_\mathrm{o}$, where the two summands only contain terms acting on even and odd bonds and commute mutually allowing for a Suzuki-Trotter decompositions of the operator
\begin{equation}
 U_t = \left[ U^\mathrm{e}_{\delta t/2} U^\mathrm{o}_{\delta  t} U^\mathrm{e}_{\delta t/2}\right]^n + \mathcal{O}(\delta t^2),
\end{equation}
where $U^\mathrm{e,o}_{t}=\exp(-i \mathcal H^\mathrm{e,o} t)$ is the time evolution operator for the even and odd bonds.
It is also possible to use more sophisticated schemes like the Forest-Ruth decomposition~\cite{omelyan02} that involves the application of seven operators but has an error of $\mathcal{O}(\delta t^5)$.
As explained in \Sref{sec:global}, the jump time distribution in the trajectory approach can actually require extremely small time steps such that higher-order expansions are not feasible in this context and we thus resort to the simpler $\mathcal{O}(\delta t^2)$ method.
In the TEBD algorithm $U^\mathrm{e}_{\delta t/2} U^\mathrm{o}_{\delta  t} U^\mathrm{e}_{\delta t/2}$ is applied at each time step by consecutive application of the two-site gates followed by an optimization of the local two-site tensors via singular value decomposition and truncation of the Schmidt spectrum.
For a detailed comparison of different time integrators such as Runge-Kutta or Krylov methods for which some can also be applied to systems with longer ranged interactions we refer to other publications such as Ref.~\cite{garciaripoll06} and references therein.

We implement the conservation of the total number of particles into our simulations for both the pure state approach as well as the superoperator algorithm where at total number of particles in the "bra"s and the "ket"s are conserved separately.
This has certain consequences for the behavior of the entanglement entropy as discussed later.


\subsection{Quantum Trajectories}
The time evolution of a mixed state $\rho$ according to a quantum master equation as given in \Eref{eq:master} can be unraveled using the quantum trajectory approach~\cite{dum92,gisin92,dalibard92,moelmer93,breuer97,plenio98,breuer10} (for a recent review see Ref.~\cite{daley14}).
The density matrix is decomposed into an average over an ensemble of pure states $\{ \ket{\Psi_i} \}$,
\begin{equation}
\rho(t) = \frac{1}{N_\mathrm{samples}} \sum_i \ketbra{ \Psi'_i(t)}{\Psi'_i(t)}
=\overline{\ketbra{ \Psi'_i(t)}{\Psi'_i(t)}},
\end{equation}
given that $\rho(0) = \ketbra{\Psi}{\Psi}$, i.e. $\ket{\Psi_i(0)} = \ket{\Psi}$.
Here, the $\ket{\Psi'(t)} = (\braket{\Psi(t)}{\Psi(t)})^{-1/2} \ket{\Psi(t)}$ simply denote the properly normalized wave functions.
The correct ensemble that will reproduce the time evolution according to \Eref{eq:master} is generated by evolving the initial wave function, that is normalized to 1, with an effective (non-unitary) Hamiltonian
\begin{equation}
\mathcal{H}_\mathrm{eff}= \mathcal{H} - i \sum_\alpha \kappa_\alpha c_\alpha^\dagger c_\alpha.
\label{eq:HamEff}
\end{equation}
As the norm at some time $\tau$ drops below a randomly drawn threshold $r \in [0,1)$, one performs a quantum jump that transforms the wave function according to
\begin{equation}
 \ket{\Psi_i(\tau)} \rightarrow \frac{c_\alpha \ket{\Psi_i(\tau)}}{ \sqrt{\bra{\Psi_i(\tau)} c_\alpha \ket{\Psi_i(\tau)}}}.
\label{eq:jump}
\end{equation}
The position $\alpha$ of the jump is chosen according to the probabilities 
\begin{equation}
p_\alpha = \frac{\bra{\Psi_i(\tau)} c_\alpha^\dagger c_\alpha \ket{\Psi_i(\tau)}}{\sum_\beta p_\beta}.
\label{eq:jumpPos}
\end{equation}

In short, the algorithm works as follows:
\begin{enumerate}[(i)]
 \item Draw a random number $r \in [0, 1)$.
 \item Evolve $\ket {\Psi_i}$ according $\mathcal{H}_\mathrm{eff}$.
 \item If the norm drops below $r$, perform quantum jump according to \Eref{eq:jump} and (\ref{eq:jumpPos}).
 \item Continue with (i).
\end{enumerate}
This scheme can be implemented using MPS~\cite{daley09,daley14} where step (ii) is performed using standard time evolution techniques (such as TEBD discussed in the previous section) and (iii) corresponds to the application of a few-site gate (in the case considered here the gate only acts on a single site).
Note that there also exist schemes that can take advantage of higher order time integrators (see Ref.~\cite{daley14} and references therein).
One has to note that the time evolution with a non-unitary Hamiltonian violates the canonical form of the MPS in \Eref{eq:MPS} and special care has to be taken to reorthogonalize the matrices.

Expectation values of some operator $\hat A$ can also be recast as an ensemble average, i.e.
\begin{equation}
\langle \hat A(t) \rangle = \mathrm{Tr}[\hat A \rho(t)] = \overline{\bra{\Psi'_i(t)} \hat A \ket{\Psi'_i(t)}}.
\end{equation}


\subsection{Superoperator Approach}
The superoperator renormalization algorithm ~\cite{zwolak04,prosen09} is based on the representation of a mixed state as a pure state in an enlarged Hilbert space.
If alike in the previous section $\lbrace \ket{i_1 i_2 ... i_L} \rbrace$ is a basis of our physical Hilbert space of local dimension $d$, the $\ket{i_1 i_2 ... i_L;j_1 j_2 ... j_L}_\#=\ketbra{i_1 i_2 ... i_L}{j_1 j_2 ... j_L}$ span the dual space of local dimension $d \times d$.
The density matrix can then be decomposed in matrix product form as
\begin{align}
\begin{split}
 \ket{\rho}_\# = \sum_{i_1 i_2 ... i_L} \sum_{j_1 j_2 ... j_L} &
\mathbf{B}_1^{i_1,j_1} \mathbf{\Lambda}_2 \mathbf{B}_2^{i_2,j_2} ...\\~ \mathbf{\Lambda}_{L} \mathbf{B}^{i_L,j_L}_L
&\ket{i_1 i_2 ... i_L;j_1 j_2 ... j_L}_\#.
\label{eq:MPSSuper}
\end{split}
\end{align}
The quantum master equation can now formally be written as a Schr\"odinger equation
\begin{equation}
\partial_t \ket{\rho}_\# = -i \mathfrak{H} \ket{\rho}_\#
\label{eq:SuperOp}
\end{equation}
with $\mathfrak{H}$ being a superoperator incorporating the right-hand site of \Eref{eq:master} -- $\mathfrak{h}$ can be thought of as a non-unitary Hamiltonian.
In analogy to the pure quantum states, \Eref{eq:SuperOp} can be integrated using the standard TEBD algorithm.
An important feature of this formulation is that thermal (Gibbs) state at temperatures can be obtained by evolving the infinitely hot initial density matrix in imaginary time with $\mathfrak{H}'=H \otimes \mathbf{1} + \mathbf{1} \otimes H$.

The only difference with respect to the pure state method is that the expectation value of some observable $\hat A$ is \textit{not} given by ${}_\#\bra \rho \hat A \ket{\rho}_\#$ but one has to directly evaluate $\mathrm{Tr}[\rho \hat A]$ given that $\mathrm{Tr}[\rho]=1$.


\subsection{Entanglement Entropy}
The MPS decomposition in \Eref{eq:MPS} is particularly useful as the $\mathbf{\Lambda}_l$ encode the eigenvalues of the reduced density matrix $\tau_i = \lambda_i^2$ when the system is bipartitioned at bond $l-1$.
Consequently, we have direct access to the entanglement entropy and the von Neumann entropy reads
\begin{equation}
S_l = -2\sum_{a=1}^\chi \lambda_a^2 \log \lambda_a.
\label{eq:entropyVN}
\end{equation}

The amount of entanglement that needs to be encoded into the Schmidt spectrum of the MPS is crucial for the matrix dimensions needed to faithfully approximate the state.
For ground-states of one-dimensional gapped systems, the area law~\cite{eisert10} ensures that the entanglement saturates for long block sizes~\cite{hastings07} and even for one-dimensional critical systems whose entanglement entropy shows a logarithmic divergence with block size can be well approximated using a polynomially increasing MPS rank~\cite{verstraete06}. 
From a practical point of view \Eref{eq:entropyVN} often is a figure of merit of how demanding in terms of resources and computer time the simulations will be.

Based on these concepts, how do we quantify entanglement for mixed states in our open system setup?
For the quantum trajectories a useful measure is to consider the entanglement entropy for each trajectory and perform an ensemble average 
\begin{equation}
	S_l= \overline{ S_l(\ket{\Psi'_i}) }. 
	\label{eq:avgEntr}
\end{equation}
as it has been used in previous works~\cite{nha04,barmettler11}.
This measure is bounded from below by the entanglement of formation, that is the minimum of $S_l$ over all decompositions of $\rho$~\cite{hill97}, but is not a good entanglement measure in the quantum information sense as it depends on the unraveling of the density matrix and is thus not unique.
The latter deficiency will be important for the performance of the trajectory method and $S_l$ should not be thought of as the entanglement of a certain mixed state but of the ensemble average for a particular unraveling of the master equation.
A second important quantity is the maximal entropy in the trajectory,
\begin{equation}
S_\mathrm{max}=\max_i \lbrace S_l(\ket{\Psi_i}) \rbrace,
\end{equation}
 ensemble as it is the actual quantity that gives us feedback about the "worst-case" entanglement for a \textit{single} trajectory that has to be accounted for in each run of the simulation. 

The MPS representation of the superoperator algorithm provides us with a natural definition of the operator space entanglement~\cite{znidaric08, pizorn09} $S_{\#,l}$.
The MPS representation in \Eref{eq:MPSSuper} provides us with a Schmidt spectrum from which $S_{\#,l}$ is defined as in \Eref{eq:entropyVN}.
For a pure state $\rho = \ketbra{\Psi}{\Psi}$ it is straightforward to see that $S_{\#,l} = 2 S_l$.
In contrast to the ensemble averages, $S_{\#,l}$ does not depend on the unraveling of the master equation and one can thus be used to identify the mixed state.

In the remainder of this paper we will restrict ourselves to the half chain bipartition, $l=L/2$, and omit the bond index.


\section{Results}
\label{sec:res}

\subsection{Global Dissipation}
\label{sec:global}

\begin{figure}[t]
\begin{center}
\includegraphics[width=\columnwidth]{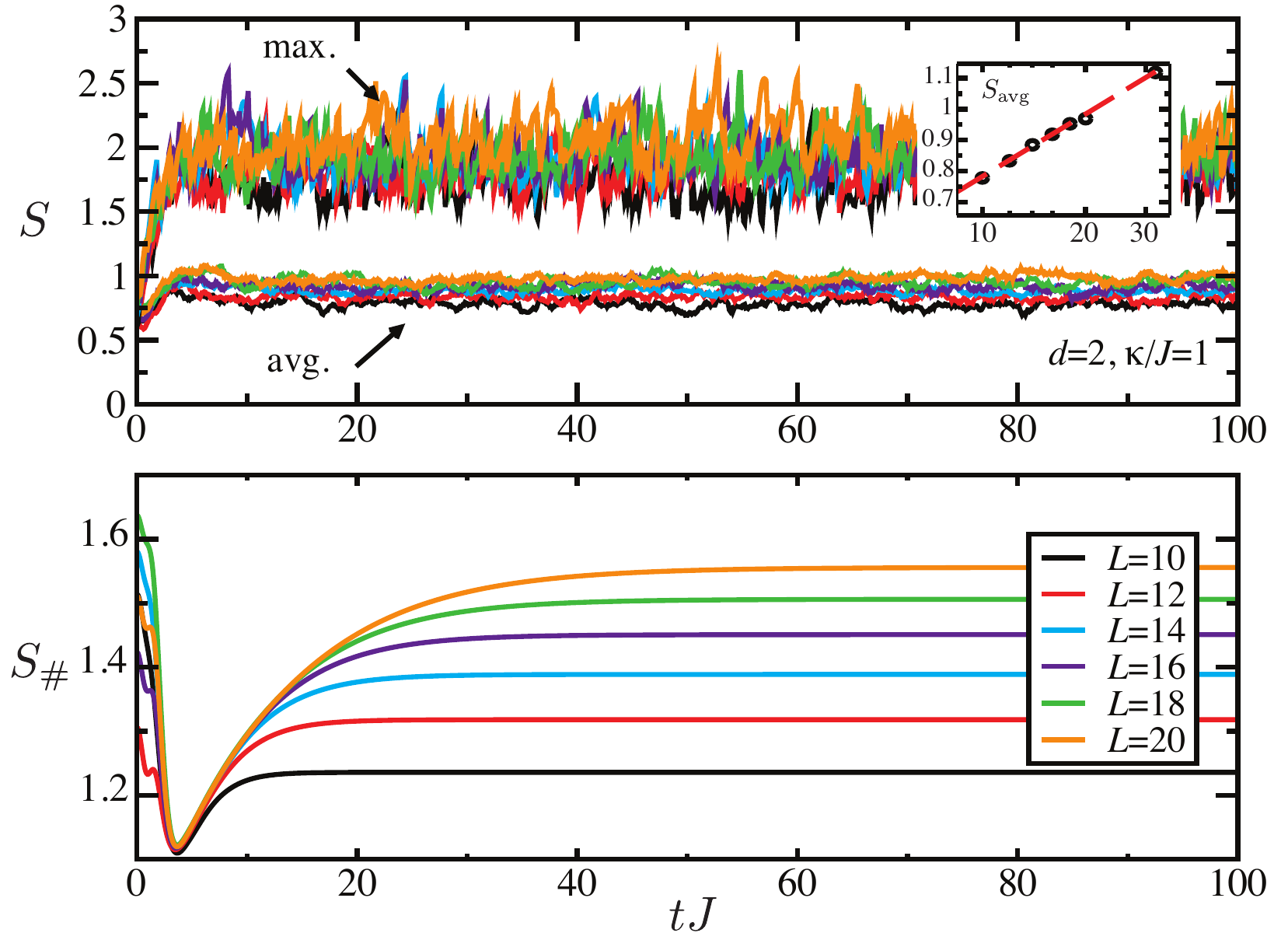}
\caption{Time evolution of the operator space entanglement $S_\#$ for global phase noise acting on hard-core bosons  at half filling for system sizes $L=10$ to 20.
The internal MPS dimension is $\chi=1000$.
\textit{Inset:} System-size scaling of the average steady-state entropy of the trajectory ensemble. The red line denotes a logarithmic fit.
}
\label{fig:Entropy_JustDM_global}
\end{center}
\end{figure}

\begin{figure}[t]
\begin{center}
\includegraphics[width=\columnwidth]{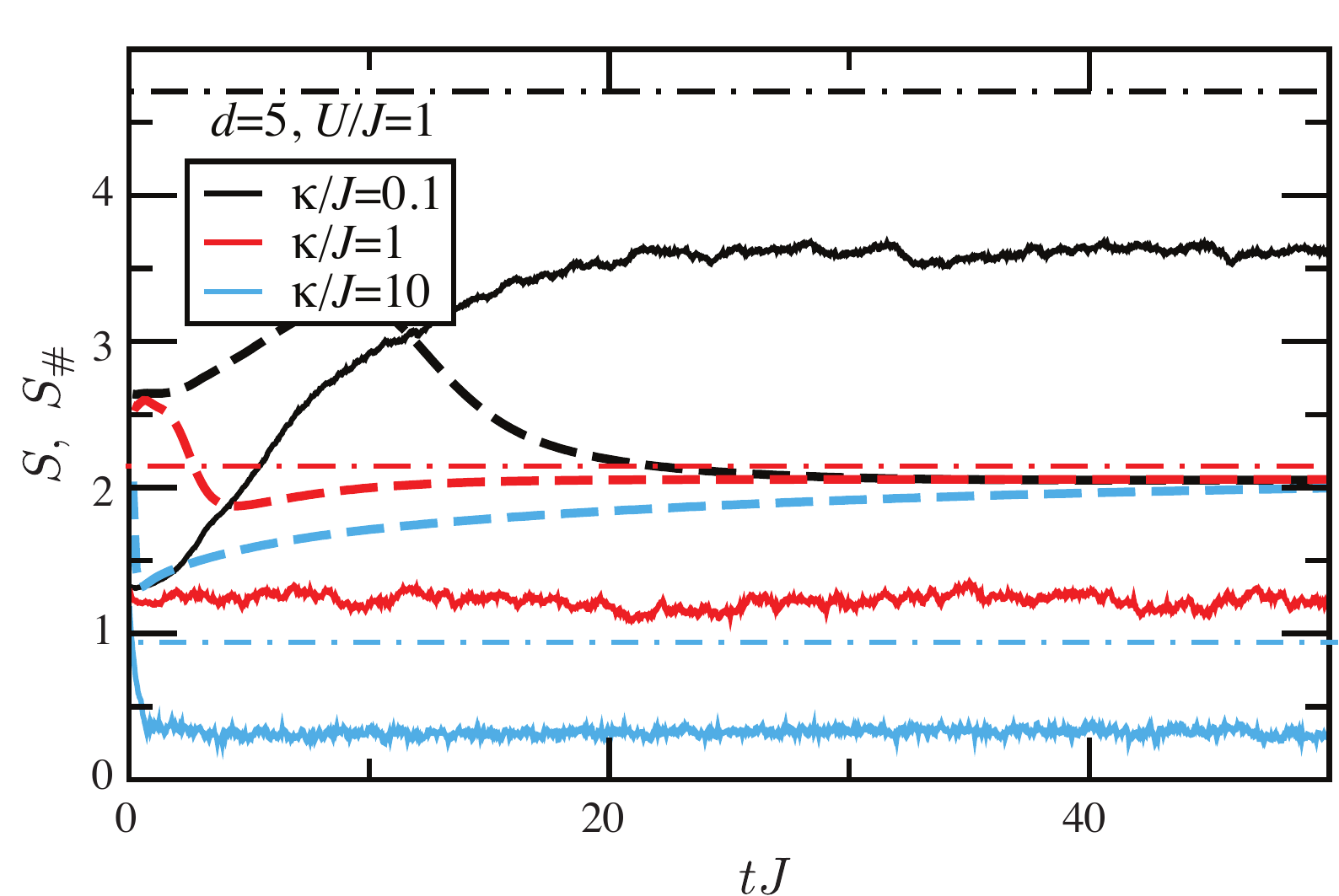}
\caption{ Entanglement entropy for global dissipation for soft-core bosons with local Hilbert space $d=5$ and , $L=10$ and $U=J$.
The full and dashed-dotted lines denote the average as well as maximal entropy from the trajectory ensemble whereas the operator space entropy of the super operator is denoted by dashed lines.
By varying $\kappa/J$ from 0.1 to 10 one can clearly see that the amount of entanglement carried by the trajectories strongly depends on the nature of the unraveling.
}
\label{fig:Entropy_JustQJ_global}
\end{center}
\end{figure}

The first case considered here is a uniform coupling of the bath degrees of freedom to the chain, i.e. $\kappa_i = \kappa$ for all $i$.
We prepare our system in the ground state of $\mathcal{H}$ at $t=0$ and then turn on the dissipation. 
We will show data for hard-core bosons at half filling and soft-core bosons at unit fillung and low $U$ so the initial ground state is superfluid with quasi off-diagonal long-range.
Thus, the initial entanglement entropy for both approaches scales like $\log L$.

\paragraph{Entropy dynamics}
The time dependence of the operator space entropy strongly depends on the interaction as well as the bath coupling.
We start by considering the hard-core case ($d=2$) at half filling fixing $\kappa=J$.
\Fref{fig:Entropy_JustDM_global} shows the different entropies introduced in the previous section as a function of time for different system sizes.
Focussing first on $S_\#$, one sees that the entropy exhibits a sharp drop to a common value followed by a slow increase until it converges to its steady state value -- the infinitely hot state.
The entropy in the final state scales logarithmically with system size which is a consequence of the global particle number conservation in the MPS algorithm that introduces trivial correlations in the system.
The Schmidt spectrum for this case be be obtained by purely combinatorial arguments~\cite{bonnes13c,bonnes14c}.
For the ensemble of trajectories, on the other hand, the average entropy seems to converge faster compared to $S_\#$ and does not resemble the features such as the entropy decrease for short times.
$S_\mathrm{avg}$ and $S_\mathrm{max}$ exhibit a large discrepancy where the former is almost constant in time whereas the maximum of the distribution of entropies increases rapidly.
The steady state value of the average trajectory entropy only increases logarithmically with system size, as shown in the inset of \Fref{fig:Entropy_JustDM_global}.

Now we turn to the soft-core case and consider a superfluid initial state with $U/J=1$ and $d=5$ which is by far more demanding computationally.
Fixing the system size, here we choose $L=10$ which is rather small but shows the important features, the entropies for different values of the dephasing rate $\kappa/J$ are shown in \Fref{fig:Entropy_JustQJ_global}.

For very large values of $\kappa/J$ the system is heated rapidly destroying all coherences in the density matrix reflected in the fast drop of both $S_\mathrm{avg}$ and $S_\#$.
The thermalization of the super operator, however, has to build-up the particle number induced correlations and thus the entropy increases to its steady state value which is independent of the $U$, $J$ or $\kappa$.
$S_\mathrm{avg}$, on the other hand, saturates very rapidly to $S_\mathrm{avg} \approx 0.4$ ($S_\mathrm{max} \approx 1$) showing that the steady state trajectory ensemble hosts only a small amount of entanglement.

The behavior for small values of $\kappa/J$ shows a qualitatively different behavior.
$S_\mathrm{avg}$ increases almost linearly for $t \lesssim 10$, comparable to a quantum quench, and saturates to a value of about 3.5 whereas the maximal entropy almost reaches values of 5 during the time evolution.
In the super operator picture $S_\#$ also exhibits a strong increase for intermediate times followed by a rapid thermalization to the infinitely hot state whose operator space entropy is much smaller compared to the amount of entropy of the trajectories.

\begin{figure}[t]
\begin{center}
\includegraphics[width=\columnwidth]{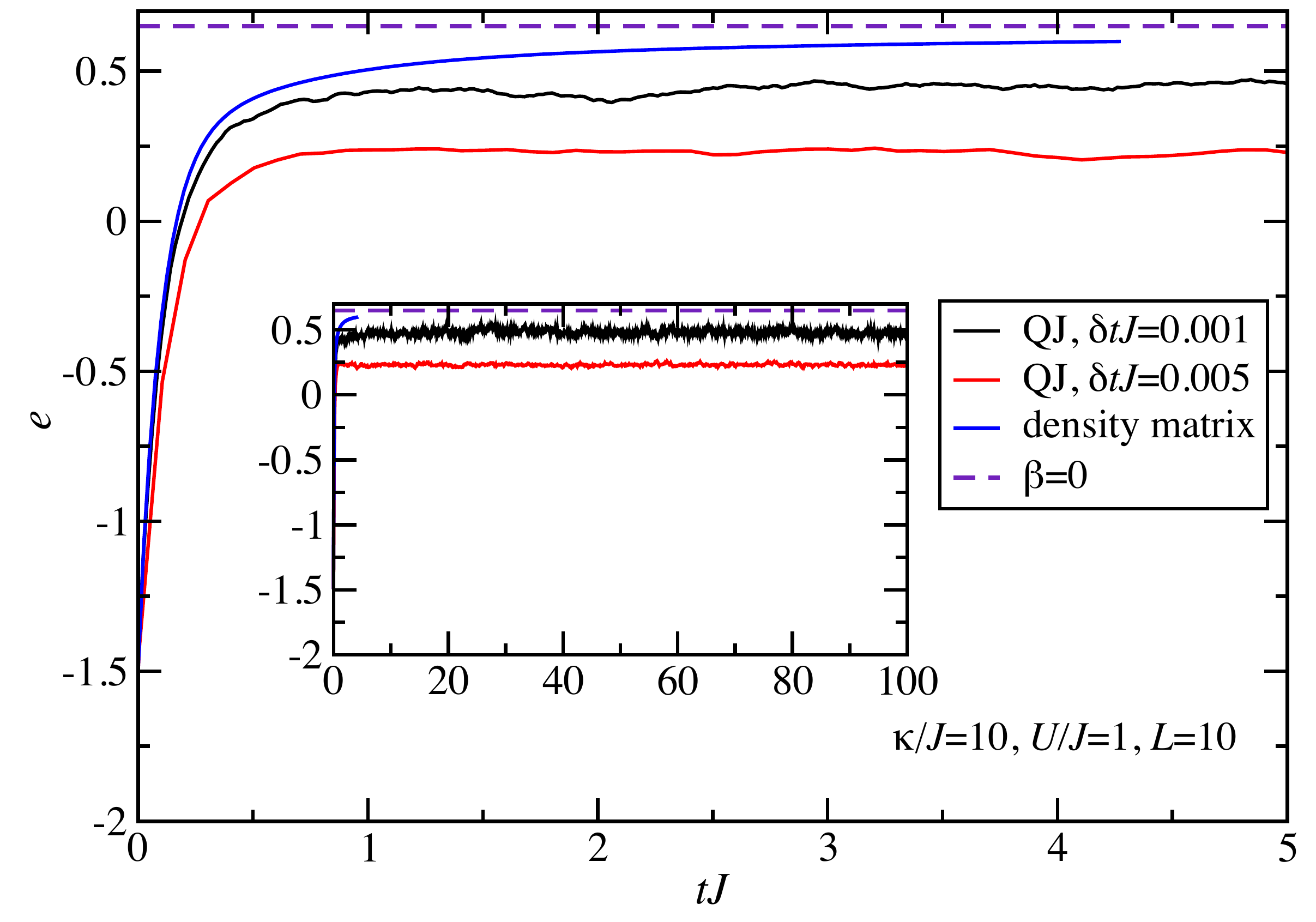}
\caption{Time dependence of the energy density $e$ for bosons at $U/J=1$ and $\kappa/J=10$ using quantum trajectories with different time steps $\delta J$ and the superoperator algorithm.
The horizontal dashed line donates the energy of the asymptotic infinite temperature state. One clearly sees how a too large Trotter time step leads to wrong results.
The inset covers the long time behavior up to $tJ=100$. 
}
\label{fig:CompareEnergies_global_kappa10}
\end{center}
\end{figure}

Strictly speaking, the von Neumann entropy do not provide a rigorous criterion of complexity of the MPS representation of a wave function but one has to consider Renyi entropies of order $\alpha<1$~\cite{schuch08}. 
A second criterion that gives a hint of how complex the MPS calculations are is the discarded weight $\epsilon=\sum_{i=\chi}^{\mathcal{D}} \lambda_i^2$ which sums the contributions to the reduced density matrix which are truncated in the optimization procedure ($\mathcal{D}$ is the dimension of the reduced density matrix).
For real-time dynamics of pure states, for instance in quantum quenches or when calculating spectral function, the entropy increase is accompanied by an increase of $\epsilon$ (when simulating at fixed $\chi$ rather than fixed $\epsilon$) eventually leading to the typical runaway phenomenon (see e.g. Ref.~\cite{gobert05}).
A similar behavior is found here both for the entropy of the trajectory ensemble as well as for $S_\#$ and thus looking at the entropies provides a good figure of merit for the required computational resources.
We like to point out that it is in fact possible to have a situation where $S_\#$ hardly increases although the truncation error shows the typical runaway phenomenon. 
Namely when performing a global quench using a thermal initial state, $\rho=\exp[-1/(k_B T) \mathcal{H}]$, one can actually have the situation where $S_\#$ shows only a slight increase in time but $\epsilon$ exhibits a rapid growth leading to a breakdown of the simulations.

What becomes evident from this discussion is that even the unraveling of the steady state, which is simply the infinitely hot density matrix, strongly depends on the dissipation rate and thus on the history of the trajectory.
In particular, small dissipation leads to a massive build up of entanglement.
In the superoperator picture, on the other hand, the final state is unique and $S_\#(t=\infty)$ is fixed.
Here the bottleneck for small $\kappa/J$ is the dynamics at intermediate times which leads to entropies of $S_\# \approx 4.5$ requiring about $\chi=1000$ states to yield converged results whereas taking about $250$ states for each trajectory suffices here.

\paragraph{Jump times}
The build-up of entanglement is rooted in the non-unitary part of the Hamiltonian whereas the quantum jumps in the case considered here remove entanglement from the system.
In order to understand the large increase of the entropy in the trajectory ensemble for small values of $\kappa$ one has to consider the number of jumps per unit time that is directly controlled by $\kappa$.
The decay of the norm during one time step $\delta t$ can be well approximated by 
\begin{equation}
\Gamma_{\Delta t}=\exp(- 2c \kappa L \Delta t),
\label{eq:Gamma}
\end{equation}
where we have assumed that $\sum_i n_i^2 = c L$ with $c$ being a constant of order of the filling.
The quantum jump operator $n_i$ tends to localize the particle at site $i$ since it performs a density measurement that counteracts the entanglement increase during the time evolution.
Thus, for large $\kappa$, the build-up of entanglement is prohibited by a large number of jumps and the state will be close to a classical one with more and more localized particles.
For small $\kappa$, on the other hand, we see a quench-like increase of $S_\mathrm{avg}$ . where the number of disentangling quantum jumps per unit time is suppressed exponentially.
The situation there is reminiscent of a global quantum quench because $\ket{\Psi_i}$ is not an eigenstate $\mathcal{H}_\mathrm{eff}$ and entanglement entropy increases linearly with time.

\begin{figure}[t]
\begin{center}
\includegraphics[width=\columnwidth]{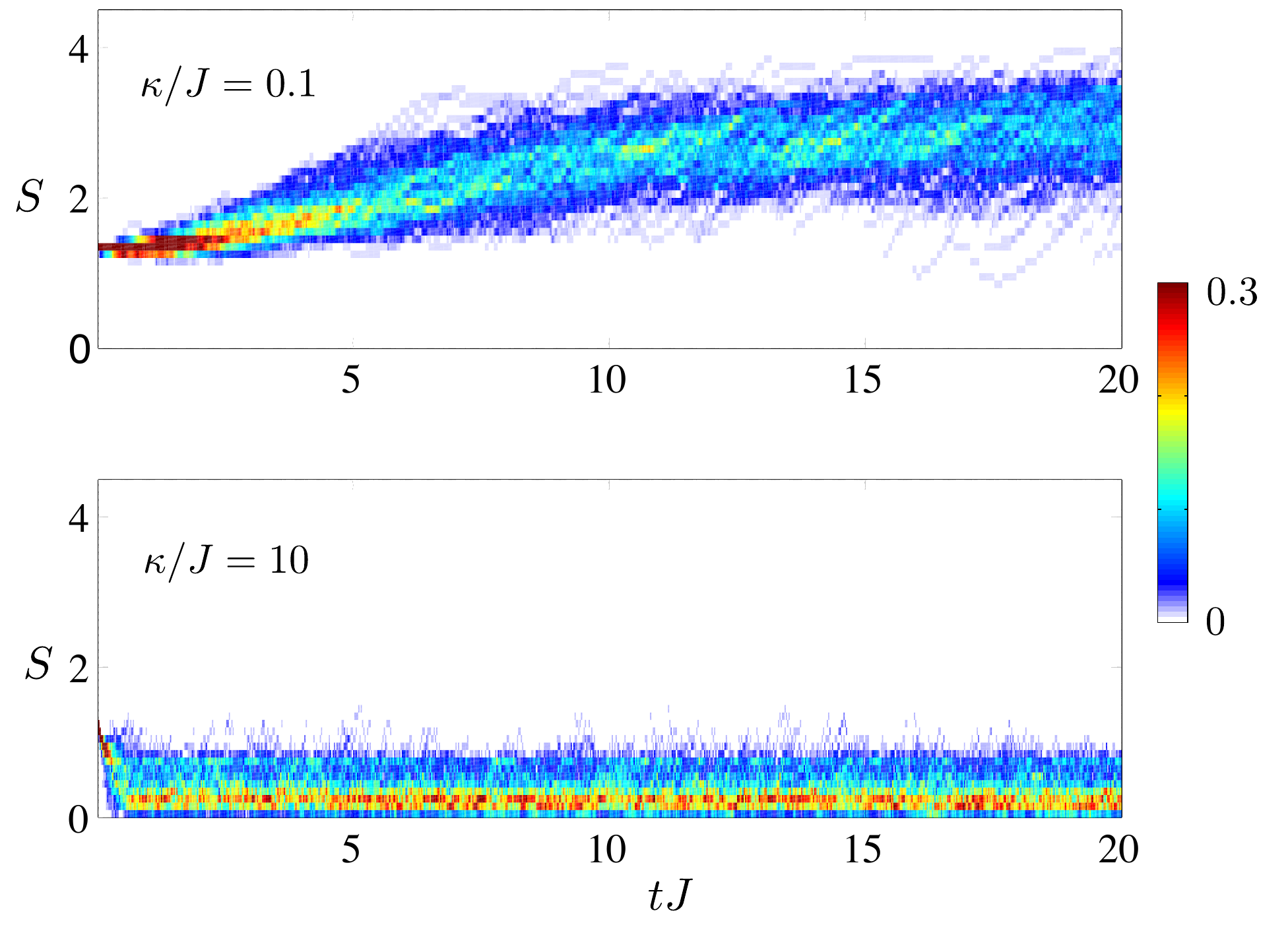}
\caption{Histogram of the entropy for a half-system bipartition of the trajectory ensemble for global dissipation with  $U=J$ and $L=10$ sites for and $\kappa/J=0.1$ (10) in the upper (lower) panel.
}
\label{fig:HistogramEntropies_Global}
\end{center}
\end{figure}

For large values of $\kappa/J$ the low entanglement in the trajectory ensemble can obviously be represented with much fewer states compared to the superoperator decomposition.
There is however an important aspect in the trajectory approach when considering global dissipation (in the sense of extensively many jump operators $c_\alpha$) since the decay rate in \Eref{eq:Gamma} is proportional to the strength of dephasing as well as the system size.
Using the approximation in \Eref{eq:Gamma} and neglecting correlations between consecutive events, we find an average time between two jumps of $\overline\tau=(2c\kappa L)^{-1}$.
A closer inspection of the jump time distribution reveals that the probability for finding a jump in a time interval $0.2\, \overline\tau$ is already above 50\% thus the time step has to be chosen over an order of magnitude smaller than the average jump time to avoid time discretization errors.
This gives an additional constraint on the choice of the time step $\Delta t$, over and above the usual time discretization error that is well understood, since the Trotter time step has to be well below $\overline \tau$.
Consequently, the number of time steps required to reach a maximal time, for instance the steady state, acquires an additional factor $L$ and one can not take advantage of higher-order methods for the time evolution.
An illustration of the time discretization error is given in \Fref{fig:CompareEnergies_global_kappa10} where the average energy for $\kappa/J=10$, $d=5$ and $U/J=1$ shows systematic deviations from the (converged) super operator result and also approaches an incorrect steady state value.

As discussed earlier there exist update schemes which are able to take advantage of higher-order time integrators which, however, do not alleviate the problem of the small jump times.

\paragraph{Distribution of Observables}

\begin{figure}[t]
\begin{center}
\includegraphics[width=\columnwidth]{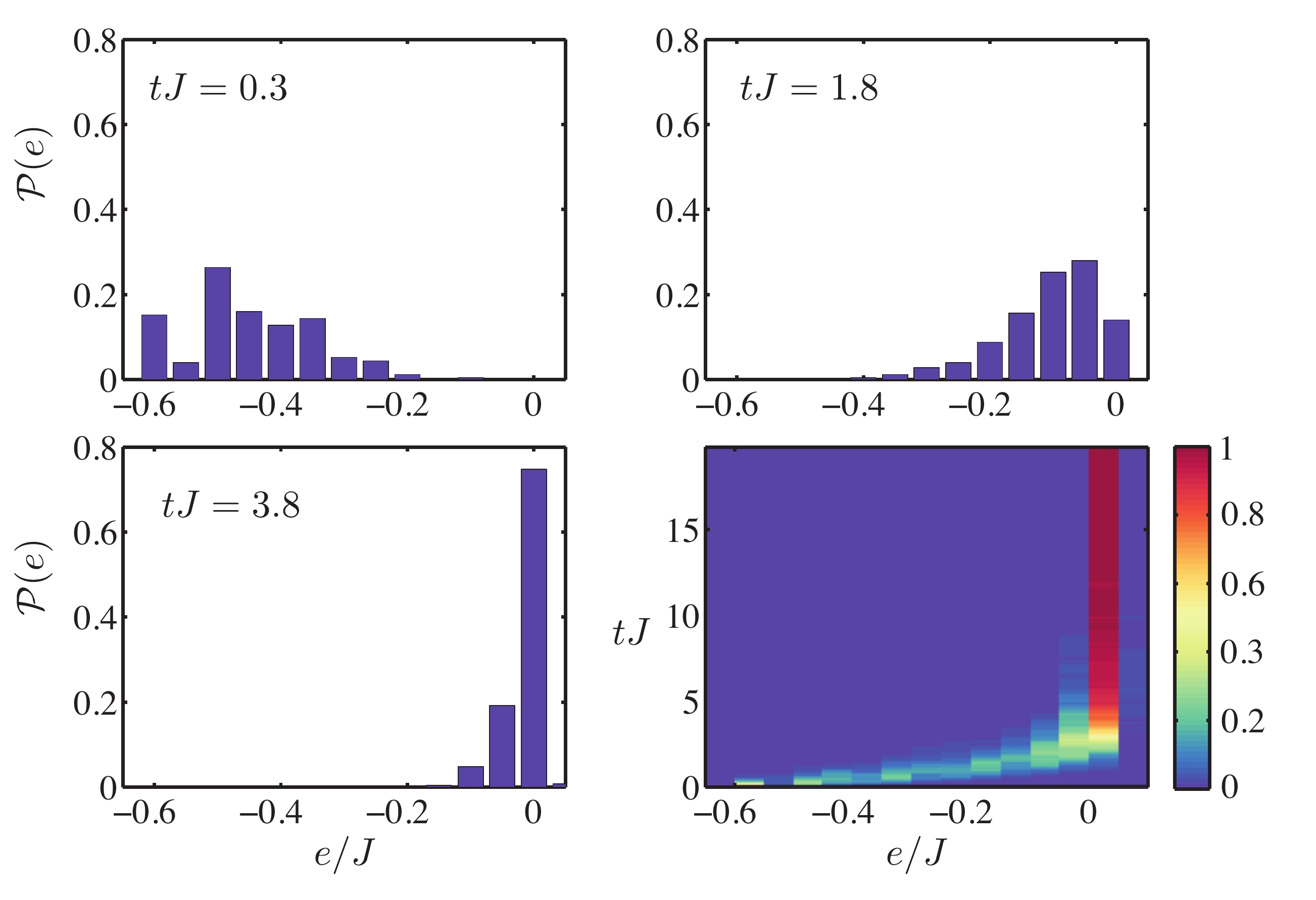}
\caption{Histograms of the energy density $e$ for the hard-core case $L=14$ with global dissipation and $\kappa=J$.  
}
\label{fig:HistogramEnergy}
\end{center}
\end{figure}

In light of these results the questions about the distribution of entanglement entropy in the trajectory ensemble rises.
\Fref{fig:HistogramEntropies_Global} shows the time evolution of the distribution for the cases discussed previously, namely $\kappa/J=0.1$ and $\kappa/J=10$.
For the former case the distribution of entanglement entropy is broadly distributed around $S_\mathrm{avg}$ indicating that the generated entanglement is not confined to some rare and highly entangled trajectories but a typical wave function is already highly entangled.
In the opposite case of small dissipation the entropy is rather peaked but also does not exhibit some features of rare events in its histogram.

The stochastic nature of the trajectory method leads to statistical uncertainties on top of the time discretization error and systematic deviations due to the truncation in the MPS optimization.
Given that the different trajectories are statistically independent (which is true as they are independent realizations of the process), the error for some observable $\langle \hat{\mathcal{O}}(t) \rangle$ can be estimated via the variance of the trajectory ensemble
\begin{equation}
\sigma_\mathcal{O}(t) = \frac{\mathrm{Var}_\mathcal{O} (\ket{\Psi_i(t)})}{\sqrt{N_\mathrm{samples}}}.
\label{eq:std}
\end{equation}
Thus, $\sigma_\mathcal{O}$ can be made arbitrarily small by increasing the number of trajectories $N_\mathrm{samples}$.
This observation, however, requires that the central limit theorem holds.
This requires that the first two moments of the distribution of $\mathcal{O}_i(t)=\langle \Psi'_i(t)| \hat{\mathcal{O}} | \Psi'_i(t) \rangle$ are well defined.

This point is not purely academic but there actually exist examples where fat tails of the distribution functions, which can appear for various reasons, can lead to problems in the Monte Carlo sampling procedure (see e.g. Refs.~\cite{dayal04,trail08,corboz08,bonnes11b}).

We therefore analyze the histograms for various observables such as the energy and the entanglement entropy and thus far we have not observed rare events or fat-tailed distributions.
For instance \Fref{fig:HistogramEnergy} shows the time evolution of distribution of the total energy density, $\mathcal{P}(e)$, which starts completely peaked at the ground state energy density and smoothly varies until it is sharply peaked around the steady state value.

\begin{figure}[t]
\begin{center}
\includegraphics[width=\columnwidth]{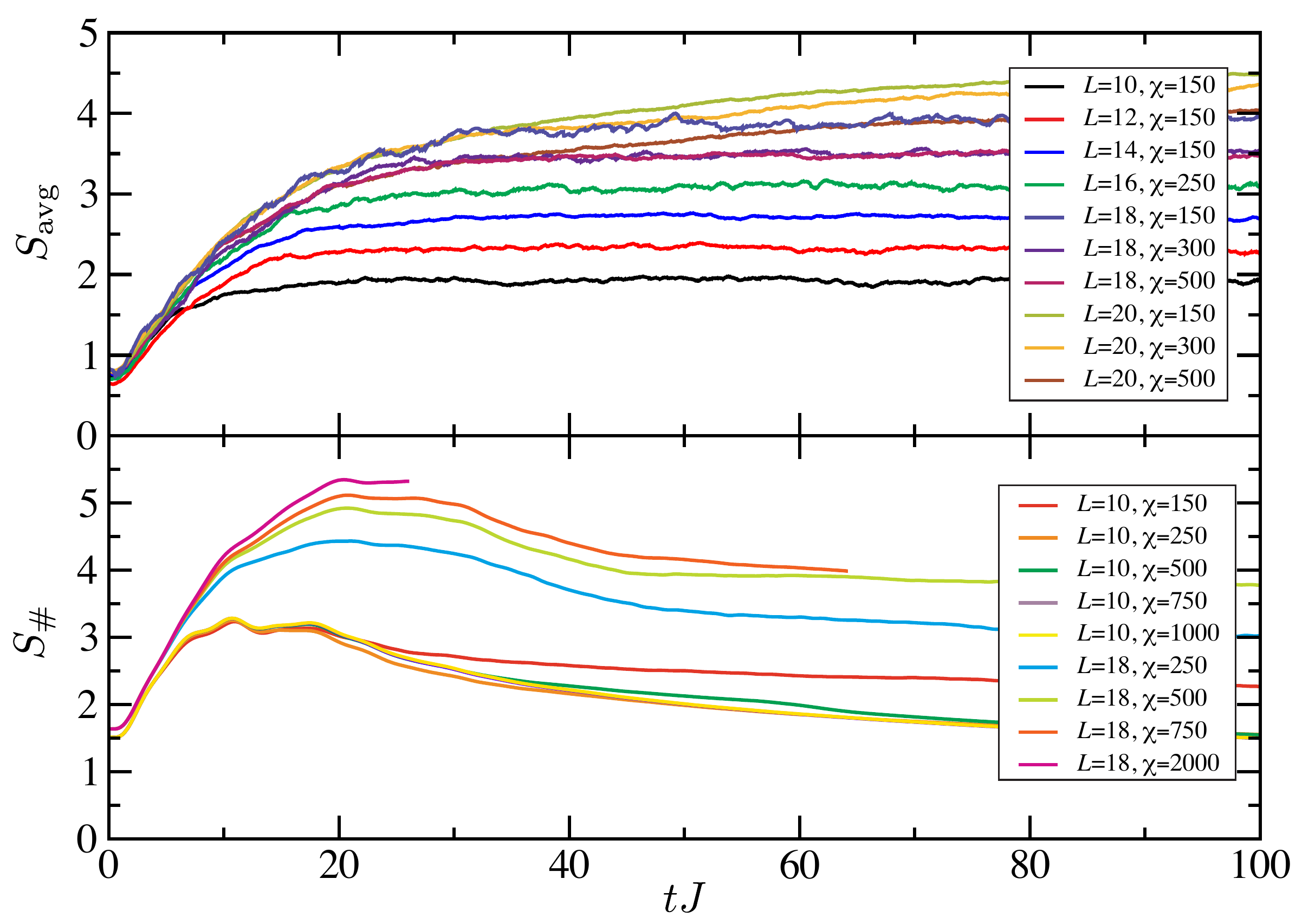}
\caption{Time evolution of the $\Smax$ (lower panel) and $\Savg$ and $S_\#$ (upper panel)  for local phase noise at $\kappa/J=1$ and $d=2$.  
}
\label{fig:Entr}
\end{center}
\end{figure}

\subsection{Local Dissipation}
\label{sec:local}
\begin{figure}[t]
\begin{center}
\includegraphics[width=\columnwidth]{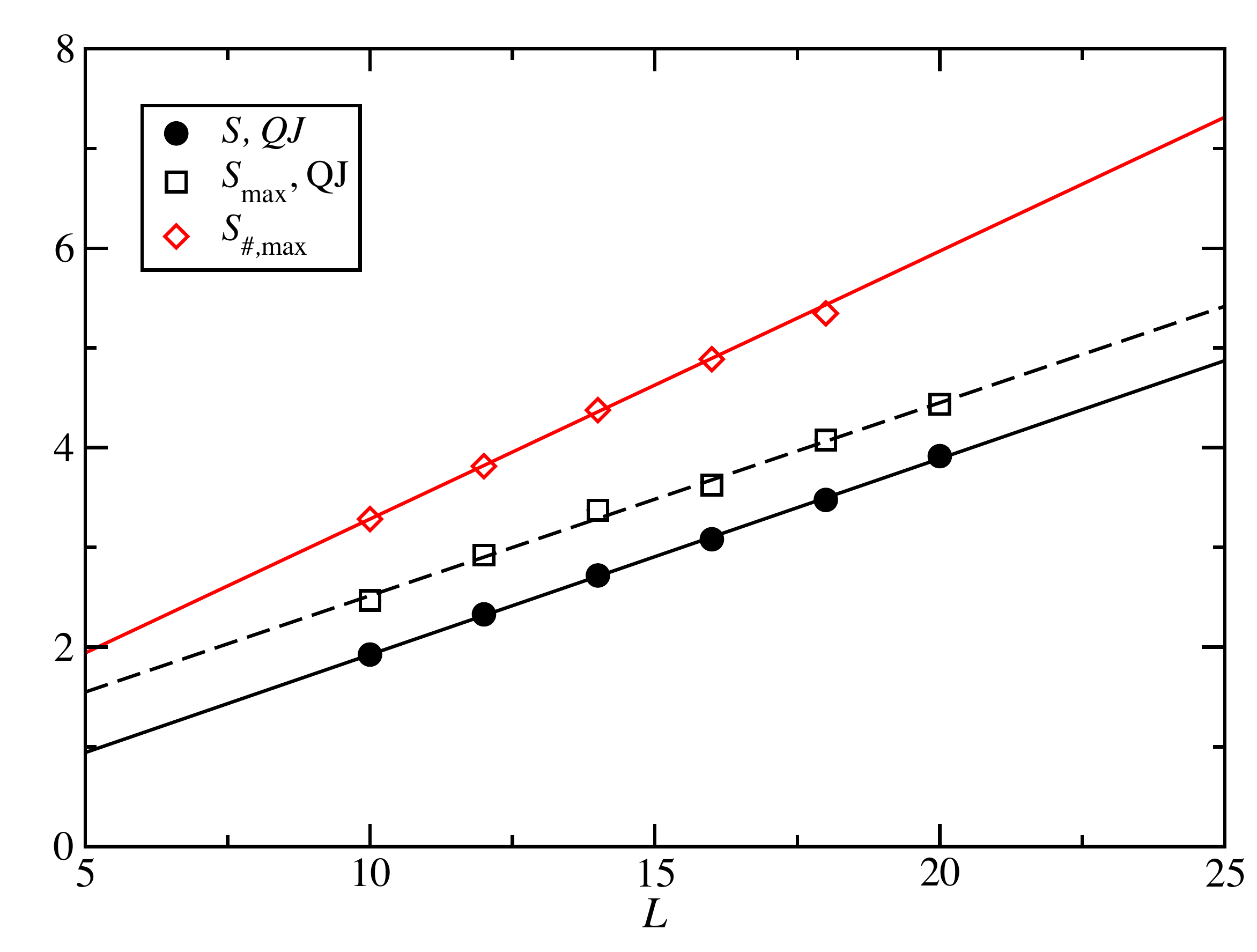}
\caption{Scaling of $\Savg$, $\Smax$ and $\max\{S\#\}$ for local dissipation using $\kappa/J=1$ and $d=2$.
The full lines are linear fits $S(L)=\alpha L + \beta$.
We find $\alpha \approx 0.25$ for $S_\#$ and $0.2$ for the $\Smax$ and $\Savg$.
}
\label{fig:ScalingEntropyLocal}
\end{center}
\end{figure}

If the bath is only coupled to a single lattice site -- here the central site at $i=L/2$ -- the behavior of the system is distinct from the results discussed in the previous section.
The average and maximal entropy of the trajectory ensemble as well as the operator space entanglement is shown in \Fref{fig:Entr} for different system sizes and MPS ranks.
One clearly sees that the entropy shows a strong scaling with system size and is greatly enhanced compared to the case of global dissipation (c.f. \Fref{fig:Entropy_JustQJ_global}).
A finite-size analysis of the maximal entropy reveals an extensive scaling of the entropy, as shown in \Fref{fig:ScalingEntropyLocal}, that severely limits the accessible system sizes.
A notable difference between $S$ and $S_\#$ is that the entropy in the trajectory ensemble increases until it reaches its (extensive) steady-state value whereas the operator space entanglement shows an extensive maximum at intermediate times -- in accordance with other studies~\cite{bonnes14c} -- but a logarithmically diverging steady-state entropy.
This can be attributed to the spreading of energy, pumped into the central site, in a quench-like scenario where quasi-particle excitations spread entanglement.
The strong increase of entanglement is also in accordance with the findings in Ref.~\cite{barmettler11}.

\begin{figure}[t]
\begin{center}
\includegraphics[width=\columnwidth]{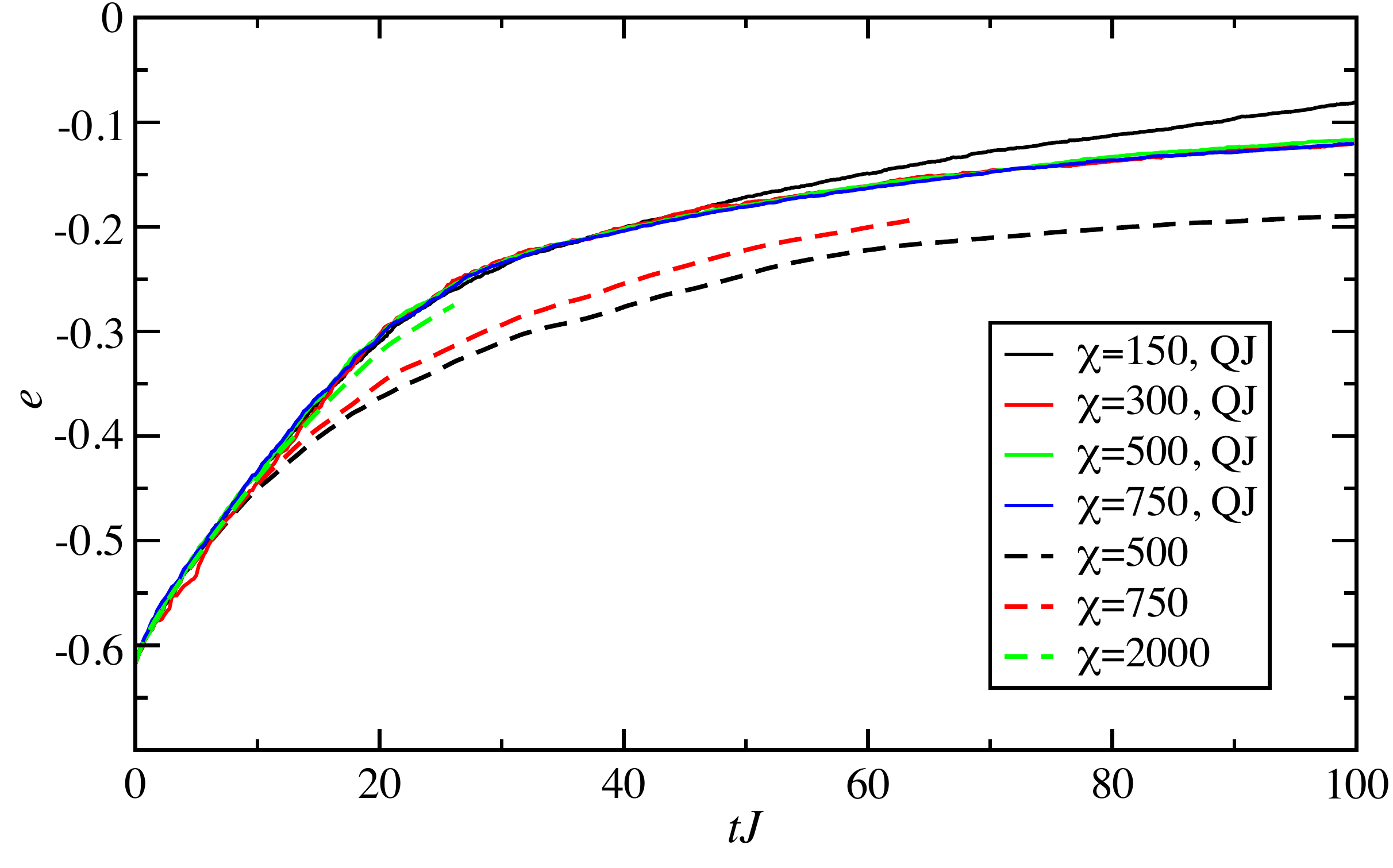}
\caption{Expectation value of the energy density $e$ for the hard-core chain with $L=10$ sites subject to local dissipation and $\kappa=J$ for different bond dimensions and methods.
}
\label{fig:EnergyQjVsSO}
\end{center}
\end{figure}

We perform a linear regression of the maximal entropies (see \Fref{fig:ScalingEntropyLocal}) and find $\alpha \approx 0.25$ for the operator space entropy compared to $\alpha \approx 0.2$ for the trajectory ensemble. 
We note that the data shown for the superoperator algorithm is not yet converged for $\chi \gtrsim 2500$ even for the small system sizes shown here.
The inability to obtain reliable results for the maximal entropy in the superoperator picture illustrates that this case is very demanding in practice.
For single trajectories, on the other hand, a moderate MPS dimension of $\chi \gtrsim 500$ seems to suffice to get relieable results.
This can also be seen looking at average quantities as the energy shown in \Fref{fig:EnergyQjVsSO}  for a chain with $L=20$ sites.
Whereas the quantum trajectory results converge for $\chi > 300$ up to times of $tJ=100$ the superoperator calculations show significant deviations even for small times of $tJ\sim 10$ for $\chi=2000$.
One interesting remark is that if $\chi$ is chosen too small $\Smax$ for the trajectory ensemble will approach the correct value from above since the system will not find the correct steady-state but entropy tends to increase even further.

\begin{figure}[t]
\begin{center}
\includegraphics[width=\columnwidth]{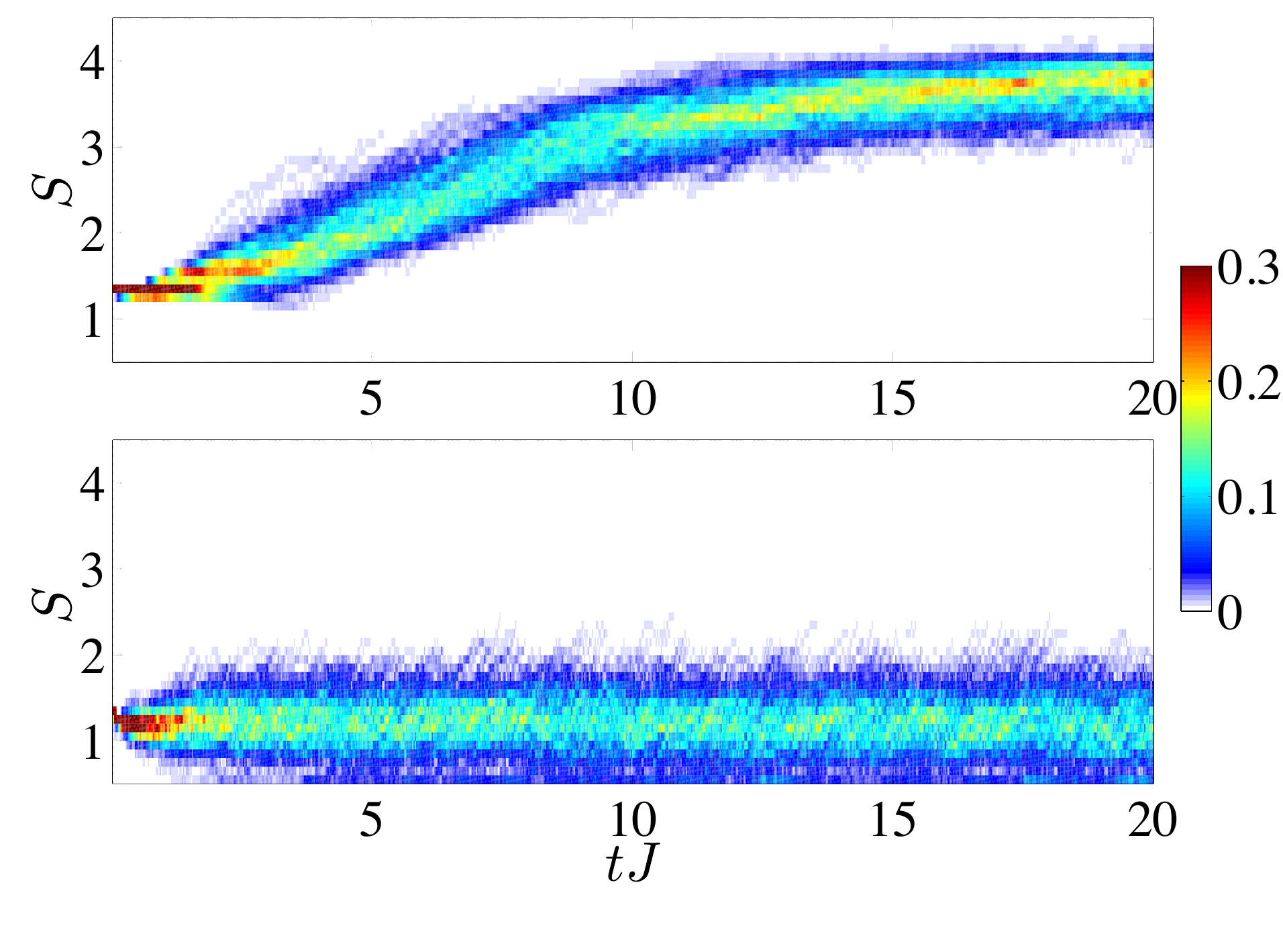}
\caption{Histogram of the entropy for a half-system bipartition of the trajectory ensemble for local dissipation for the hard-core case, and $\kappa/J=10$ (upper panel) and $\kappa/J=0.1$ (lower panel).
}
\label{fig:HistogramEntropies}
\end{center}
\end{figure}

Finally, \Fref{fig:HistogramEntropies} shows the time evolution of the entropy distribution in the trajectory ensemble. 
Alike in the previous case of global dissipation, it is strongly peaked around its mean value and we do not find evidence for rare highly entangled states that dominate $\Smax$.

\section{Comparison}
\label{sec:discussion}
Having analyzed the entanglement properties as well as the distribution of the expectation values for the trajectory ensembles, the question remains: Which methods performs better?

Let us first address the case of global dissipation considered in \Sref{sec:global}.
On the one hand, the superoperator method is plagued by a large local Hilbert space that is as large as $d^2=25$ for the cases considered here.
And even for comparable values of the entanglement the internal dimension of the MPS has to be much larger for the superstate than for the conventional MPS. 
For example, take again the picture of representing a pure state as a density matrix.
The Schmidt spectrum of $\rho$ is given by $\lambda_i \lambda_j$, where $\{ \lambda_i \}$ is the Schmidt spectrum of the pure state.
However, one can exploit two good quantum numbers associated to the fillings of the "bra"- and "ket"-index of the superstates as well as hermiticity ($\rho = \rho^\dagger$) such that the naive $d^6$ scaling is allayed.

Typically, one requires a few hundred to thousands of trajectories -- a process that can be parallelized trivially -- to get precise results for the transient dynamics and for low $\kappa$, the build-up of entanglement is found to be severe for the single trajectories such that is is more favorable to use a superoperator with adaptive $\chi$ and high-order Trotter expansion since one only has to overcome the transient entropy maximum and then relaxes into the low entangled steady state.
For large $\kappa$, the time step necessary to sample the jump events correctly becomes very small but, on the other hand, the entropy is extremely small.
From our simulations, the latter effect outweighs the small time step size due to the $\chi^3$ scaling the the TEBD algorithm.
If one is only interested in steady-state properties, the trajectory approach also offers the possibility to perform long-time averages on few trajectories hence reducing the computational effort.
For large $\kappa$, on the other hand, the small time step limits the applicability of the trajectory method for long-time results severely.

Having access to the operator space entropy can also give further physical insight since it is not "history depended" as for the trajectories e.g., one can see directly within the superoperator picture that the entropy of the steady-state reaches the same value independent of $U$ or $\kappa$ whereas the entanglement of the single trajectories has no direct physical insight into the properties of the mixed state although the outcomes of single trajectories can give interesting insight into the real time dynamics of the system also in view of experimental realizations.

The situation is different for local dissipation discussed in \Sref{sec:local}. 
Although both methods suffer from an extensive scaling of the entropy, the system sizes available with the superoperator technique are limited since a few thousand states have to be kept to the capture the transient behavior of the density matrix.
Quantum trajectories allow us to explore larger system although they are also limited by the scaling of the entropy.

One important advantage of the superoperator approach is the lack of statistical uncertainties.
Although the errors can be made arbitrarily small, operations such a Fourier transformations require a careful analysis and good quality of data.

\section{Conclusion}
\label{sec:conclusion}
We presented a comparison of the MPS implementation of the superoperator renormalization group algorithm and the quantum trajectory method for open quantum systems studying a Bose chain with dephasing.
Whereas the quantum trajectory method for global dissipation is plagued by extremely short jump times, the entropy increase in the trajectory ensemble is comparable to the operator space entropy for not too small values of the dissipation.
The trajectory approach seems particularly well suited to calculate global quantities and long-time averages for steady states whereas the superoperator approach is particularly useful for small dissipation rates and short-time behavior.
Local dissipation, on the other hand, leads to extensive entropy scaling during the time evolution limiting both approaches due to the large amount of entanglement generated.
We find however that in this case the trajectory approach allows us to access slightly larger system sizes and can in fact outperform the superoperator ansatz in this extreme case.

Having presented data for a specific problem, an import question is how generic these results are.
First of all, the superoperator provides a framework to choose a wider class of initial state that can be, in particular, thermal as it has been used in a recent study for a model of engineered dissipation~\cite{bonnes14a}.
Thermal initial states might also be implemented into the quantum trajectory method using METTS~\cite{white09,stoudenmire10,bonnes14c}, for instance, but the demonstration of the applicability and performance of such approaches are still outstanding.
An open problem is also the application to two-dimensional tensor network methods.

\section*{Acknowledgments}
We thank A. J. Daley and C. Kollath for discussions and shared insights
and acknowledge discussions with B. Kraus as well as H. Ritsch and W. Niedenzu.
This work was supported by the Austrian Ministry of Science BMWF as
part of the UniInfrastrukturprogramm of the Forschungsplattform
Scientific Computing at LFU Innsbruck, by the FWF SFB.

\bibliographystyle{apsrev4-1}

%

\end{document}